\documentclass[aps,prl,twocolumn,superscriptaddress]{revtex4}
\usepackage{graphicx}
\usepackage{color}

\newcommand{\hpr}[2]{\ensuremath{\langle #1| #2\rangle}}
\newcommand{\ket}[1]{\ensuremath{| #1 \rangle}}

\begin{document}

\title{Manifestation of nonlocal electron-electron interaction in graphene}
\author{S\o ren Ulstrup}
\affiliation{Department of Physics and Astronomy, Interdisciplinary Nanoscience Center (iNANO), Aarhus University, Denmark}
\author{Malte Sch\"uler}
\affiliation{Institut f\"ur Theoretische Physik, Unversit\"at Bremen, Otto-Hahn-Allee 1, 28359 Bremen, Germany}
\affiliation{Bremen Center for Computational Materials Science, Unversit\"at Bremen, Am Fallturm 1a, 28359 Bremen, Germany}
\author{Marco Bianchi}
\affiliation{Department of Physics and Astronomy, Interdisciplinary Nanoscience Center (iNANO), Aarhus University, Denmark}
\author{Felix Fromm}
\affiliation{Lehrstuhl f{\"u}r Technische Physik, Universit{\"a}t Erlangen-N{\"u}rnberg, Germany}
\affiliation{Institut f{\"u}r Physik, Technische Universit{\"a}t Chemnitz, Germany}
\author{Christian Raidel}
\affiliation{Lehrstuhl f{\"u}r Technische Physik, Universit{\"a}t Erlangen-N{\"u}rnberg, Germany}
\affiliation{Institut f{\"u}r Physik, Technische Universit{\"a}t Chemnitz, Germany}
\author{Thomas Seyller}
\affiliation{Institut f{\"u}r Physik, Technische Universit{\"a}t Chemnitz, Germany}
\author{Tim Wehling}
\affiliation{Institut f\"ur Theoretische Physik, Unversit\"at Bremen, Otto-Hahn-Allee 1, 28359 Bremen, Germany}
\affiliation{Bremen Center for Computational Materials Science, Unversit\"at Bremen, Am Fallturm 1a, 28359 Bremen, Germany}
\author{Philip Hofmann}
\affiliation{Department of Physics and Astronomy, Interdisciplinary Nanoscience Center (iNANO), Aarhus University, Denmark}
\affiliation{email: philip@phys.au.dk}
\date{\today}
\begin{abstract}
Graphene is an ideal platform to study many-body effects due to its semimetallic character and the possibility to dope it over a wide range. Here we study the width of graphene's occupied $\pi$-band as a function of doping using angle-resolved photoemission. Upon increasing electron doping, we observe the expected shift of the band to higher binding energies. However, this shift is not rigid and the bottom of the band moves less than the Dirac point. We show that the observed shift cannot be accounted for by band structure calculations in the local density approximation but that non-local exchange interactions must be taken into account.
\end{abstract}
\maketitle

Many-body interactions are the key to a wide range of phenomena in solids including  magnetism, superconductivity and other interesting ground states, and their understanding has long been a central objective of condensed matter physics. While the interactions do not change the volume of a metal's Fermi surface \cite{Luttinger:1960}, they lead to a modification of the occupied band width or, in the case of semiconductors, of the band gap. This has been widely used to access the self-energy in materials in order to aid the theoretical understanding of exchange and correlation effects \cite{Louie:1992}. A key-parameter controlling these is the electron (or hole) density. This can easily be changed in a semiconductor \cite{Kalt:1992aa,Trankle:1987aa,Das-Sarma:1990aa} but not in a metal.  Here we use graphene, the notable exception to this rule, to experimentally vary the electron density over a wide range. We compare the resulting band width changes to different types of electronic structure calculations.  This gives an unprecedented insight into previously inaccessible parameter regimes where the interplay of doping and non-local exchange interactions largely shapes the electronic structure.

A direct and accurate way to access many-body effects in metals is via their effect on the band structure, as observed by angle-resolved photoemission (ARPES). The many-body effects give rise to an electronic self-energy that causes the band structure to be renormalized, i.e. to deviate from a single-particle calculation. The renormalization of the total occupied band width reveals the importance of many-body effects in even the simplest situations, such as in the free electron metal sodium \cite{Jensen:1985,mcclain_spectral_2015}. The comparison between the experimentally determined electronic structure of materials and calculations that include the many-body effects using different approximations has therefore been an important tool to develop an understanding of exchange and correlation effects in solids \cite{Hedin:1965,Jensen:1985,Northrup:1989,Mahan:1989,Zhu:1991,Louie:1992,Mahan:2000aa,mcclain_spectral_2015}. Similar approaches can be used to study the effect of excited carriers on the band gap in semiconducting materials, revealing effects such as negative electronic compressibility \cite{He:2015ab,Riley:2015aa} or giant band gap renormalizations \cite{Steinhoff_NanoLett2014}. In such situations, however, it is not possible to determine the size of the band gap in the un-doped case from ARPES alone. 

\begin{figure*}
\includegraphics[width=0.7\textwidth]{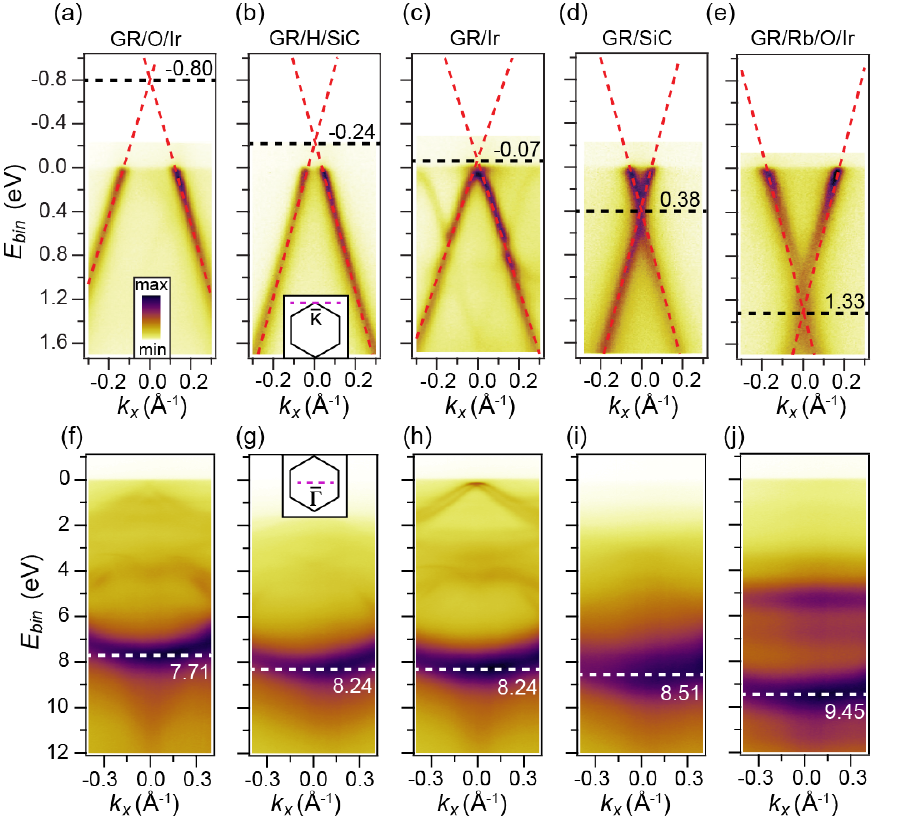}\\
\caption{(Color online) Angle-resolved photoemission data used to determining the doping and $\pi$-band width for five different epitaxial graphene systems: oxygen-intercalated graphene on Ir(111) (GR/O/Ir) \cite{Larciprete:2012}, hydrogen-intercalated graphene on SiC (GR/H/SiC) \cite{Bostwick:2010,Speck:2011,Johannsen:2013}, graphene on Ir(111) (GR/Ir) \cite{Pletikosic:2009,Larciprete:2012}, graphene on SiC (GR/SiC) \cite{Bostwick:2007} and rubidium-doped, oxygen intercalated graphene on Ir(111) (GR/Rb/O/Ir) \cite{Ulstrup:2014e}. (a)-(e) Dispersion near $\bar{K}$, showing how the Dirac point energy $E_D$ is obtained by extrapolation of the $\pi$-band (red dashed lines). (f)-(j) Corresponding dispersion near the bottom of the band at $\bar{\Gamma}$ where the position is obtained by a simple peak fitting. The inset in (a) shows the color scale used and the insets in (b) and (g) give the scan directions used for the dispersion near $E_D$ and near $E_{\pi}$, respectively.}
  \label{fig:1}
\end{figure*}

Unlike in a normal metal, the filling of the $\pi$-band in graphene can be varied by doping such that the electron density-dependence of many-body effects can be explored in the same material \cite{Bostwick:2007,Castro-Neto:2009,Elias:2011,Siegel:2011,Kotov:2012,Siegel:2013}. Near the Fermi level, several many-body effects, such as the electron-electron, electron-phonon and electron-plasmon interaction, are simultaneously present and, since they can all be doping dependent, they are difficult to disentangle  \cite{Bostwick:2007,Siegel:2011}. Here we study the electron-electron interaction in graphene in a similar way as in the pioneering work on sodium  \cite{Jensen:1985}, by measuring the width of graphene's $\pi$-band as a function of electron / hole filling and we compare the results to calculations incorporating different levels of many body interactions. We show that agreement with the experimental data can only be achieved by taking non-local exchange effects into account. 

We combine data of differently doped epitaxial graphene systems: oxygen-intercalated graphene on Ir(111) (GR/O/Ir) \cite{Larciprete:2012}, hydrogen-intercalated graphene on SiC (GR/H/SiC) \cite{Bostwick:2010,Speck:2010aa,Speck:2011,Johannsen:2013}, graphene on Ir(111) (GR/Ir) \cite{Pletikosic:2009,Larciprete:2012}, graphene on SiC (GR/SiC) \cite{Bostwick:2007,Emtsev:2009,Ostler:2010} and rubidium-doped, oxygen-intercalated graphene on Ir(111) (GR/Rb/O/Ir) \cite{Ulstrup:2014e}. All data were collected on the SGM-3 beamline of ASTRID2 \cite{Hoffmann:2004} using a photon energy of 47~eV. The sample temperature was $\approx$80~K and the total energy and angular resolution were better than 20~meV and 0.2$^{\circ}$, respectively. For the details of the sample preparation, refer to the references given in connection with the list of samples above. Note that all results shown here are slices extracted from data taken in two-dimensional $k$-space ($k_x$, $k_y$), so that cuts through high symmetry directions could be determined unambiguously.

Figure \ref{fig:1} gives the experimental results of this investigation. Panels (a)-(e) show the dispersion of the $\pi$-band near the $\bar{K}$-point of the Brillouin zone for the different graphene systems mentioned above. The dispersion is measured along a line perpendicular to the $\bar{\Gamma}-\bar{K}$ direction, such that both branches of the Dirac cone can be observed. Depending on the system, very different doping regimes can be reached, ranging from strongly $p$-doped GR/O/Ir to strongly $n$-doped GR/Rb/O/Ir. The linear dispersion of the $\pi$-band with the crossing at the Dirac point energy $E_D$ is illustrated by the red dashed lines.  $E_D$ is determined by an extrapolation of the $\pi$-bands over a wide energy range to avoid complications due to band renormalization caused by electron-phonon coupling close to the Fermi energy. In the $n$-doped case and for GR/Ir, the crossing of the extrapolated band obviously agrees very well with the observed Dirac point position. Fig. \ref{fig:1}(f)-(j) show the corresponding data for the bottom of the $\pi$-band, reaching the energy $E_{\pi}$  at the $\bar{\Gamma}$ point of the Brillouin zone. The determination of $E_{\pi}$ is even more straight-forward because this energy can be obtained by fitting an energy distribution curve through the data obtained at normal emission.  

\begin{figure}
\includegraphics[width=0.4\textwidth]{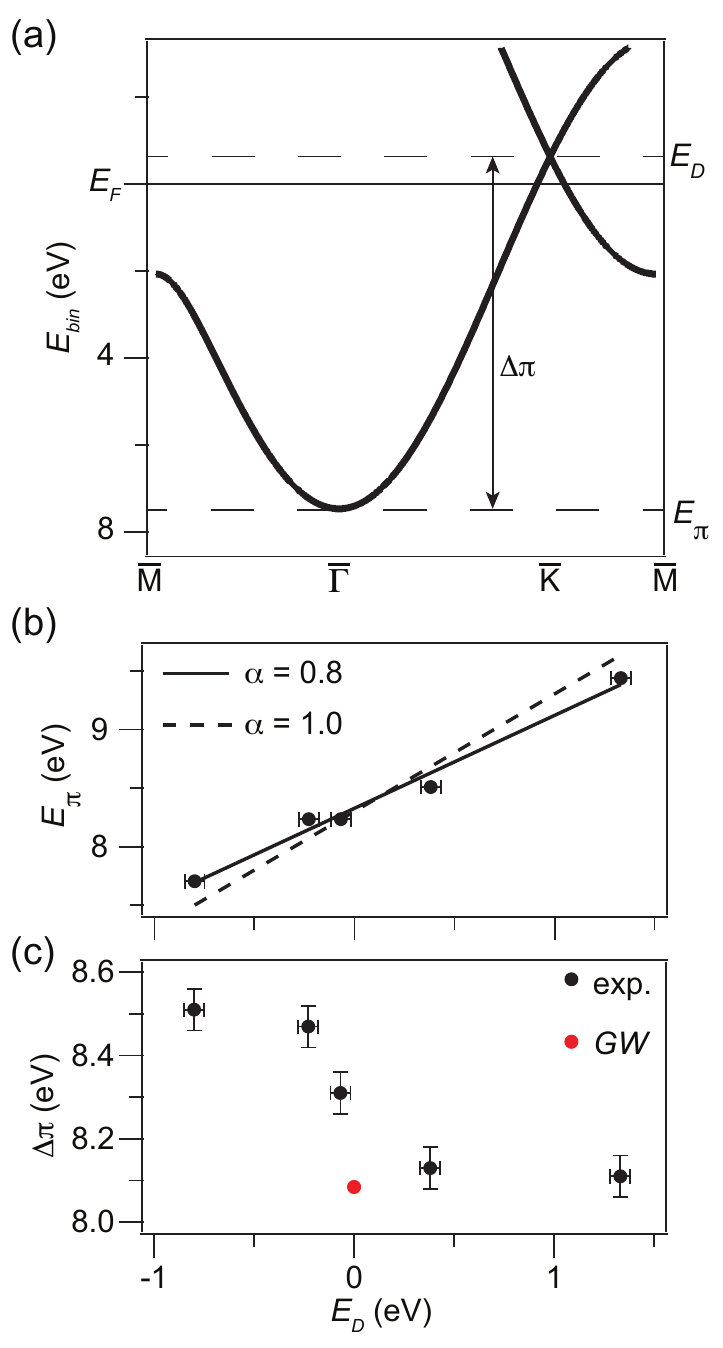}\\
\caption{(Color online) (a) Schematic band structure for the $\pi$-band of graphene defining the Dirac point energy / doping level $E_D$, the band bottom energy / occupied band width $E_{\pi}$ and the total $\pi$-band width $\Delta \pi$. (b) Occupied band width $E_{\pi}$ as a function of doping level $E_D$. The solid line is a linear fit to the data points. The dashed line is a fit with the constraint $\alpha = 1$, where $\alpha$ is the slope of the line. (c) Total band width $\Delta \pi$ as a function of  $E_D$. The red dot in (c) is the result of a $GW$ calculation for charge-neutral graphene \cite{Trevisanutto:2008}.}
  \label{fig:2}
\end{figure}

A first inspection of the shifts of $E_D$ and $E_{\pi}$ in Fig. \ref{fig:1} shows the expected trend: Doping more electrons into the $\pi$-band shifts all the states to higher binding energy. A quantitative analysis, however, reveals that the situation does not correspond to the rigid shift expected in a single-particle picture. The relevant parameters for this analysis are defined in the sketch of the $\pi$-band in Fig. \ref{fig:2}(a). The $\pi$-band as such is characterized by the parameters already introduced: the Dirac point with energy $E_D$  at $\bar{K}$ and the highest binding energy $E_{\pi}$ at $\bar{\Gamma}$, which corresponds to the occupied band width. We define the \emph{total} width of the $\pi$-band as the energy difference $\Delta \pi = |E_D - E_{\pi}|$. In addition to this, the graphene may be doped, such that $E_D$ does not necessarily coincide with the Fermi energy $E_F=0$. In the sketch, the situation for $p$-doped graphene is shown. The level of doping is directly given by $E_D$.

For a quantitative analysis of the data, we first plot $E_{\pi}$ as a function of the Dirac point energy $E_D$ in Fig. \ref{fig:2}(b). For a rigid band shift upon doping, $E_{\pi}$ should simply track the movement of $E_D$. A linear fit (solid line) does  describe the data well but the slope $\alpha$ of the line significantly deviates from 1 (see dashed line representing $\alpha=1$), indicating a many-body effect-induced band deformation upon doping. Indeed, this simple test illustrates the advantages of using graphene for experimentally probing many-body effects, as their importance is immediately seen in such raw data. This is in sharp contrast to the situation in a normal metal, where the manifestation of many-body effects only emerges via comparison to a calculated band width.  Note that a non-rigid deformation of the $\pi$-band upon doping has also been observed by Bostwick \emph{et al.} who found  the tight-binding parameters needed to describe the $\pi$-band to be doping-dependent \cite{Bostwick:2007b}.  

To quantitatively account for the many-body effects and for comparison to calculations, we plot $\Delta \pi$ as a function of $E_D$ in Fig. \ref{fig:2}(c). In the absence of many-body effects, $\Delta \pi$ would be expected to be independent of $E_D$ but it is observed to decrease with an increasing $E_D$, corresponding to the deviation of the fit's slope from unity in Fig. \ref{fig:2}(b). This observation gives an important hint as to the functional form of the many-body effects: The self-energy corrections must be odd with respect to the change from electron doping to hole doping. 

Fig. \ref{fig:2}(c) also shows the $\pi$-band width from a $GW$ calculation for charge-neutral graphene \cite{Trevisanutto:2008} which fits relatively well with the observed experimental value.  This calculated band width is somewhat ($\approx$ 6\%) larger than for a density functional theory (DFT) calculation, in agreement with the trend observed for graphite \cite{Heske:1999}. Note, however, that the experimental results for graphene near charge-neutrality can also be expected to suffer most from systematic error due to substrate screening, an effect that is not likely to be significant at the very high carrier densities in the end of the doping range investigated here (between $\approx 5\times 10^{13}$ holes and $\approx 1.3\times 10^{14}$ electrons per cm$^2$). On the other hand, it is observed that the data point for nearly neutral graphene from the GR/Ir system fits well into the smooth trend observed for the whole data set.

\begin{figure}
 \includegraphics[width=0.49\textwidth]{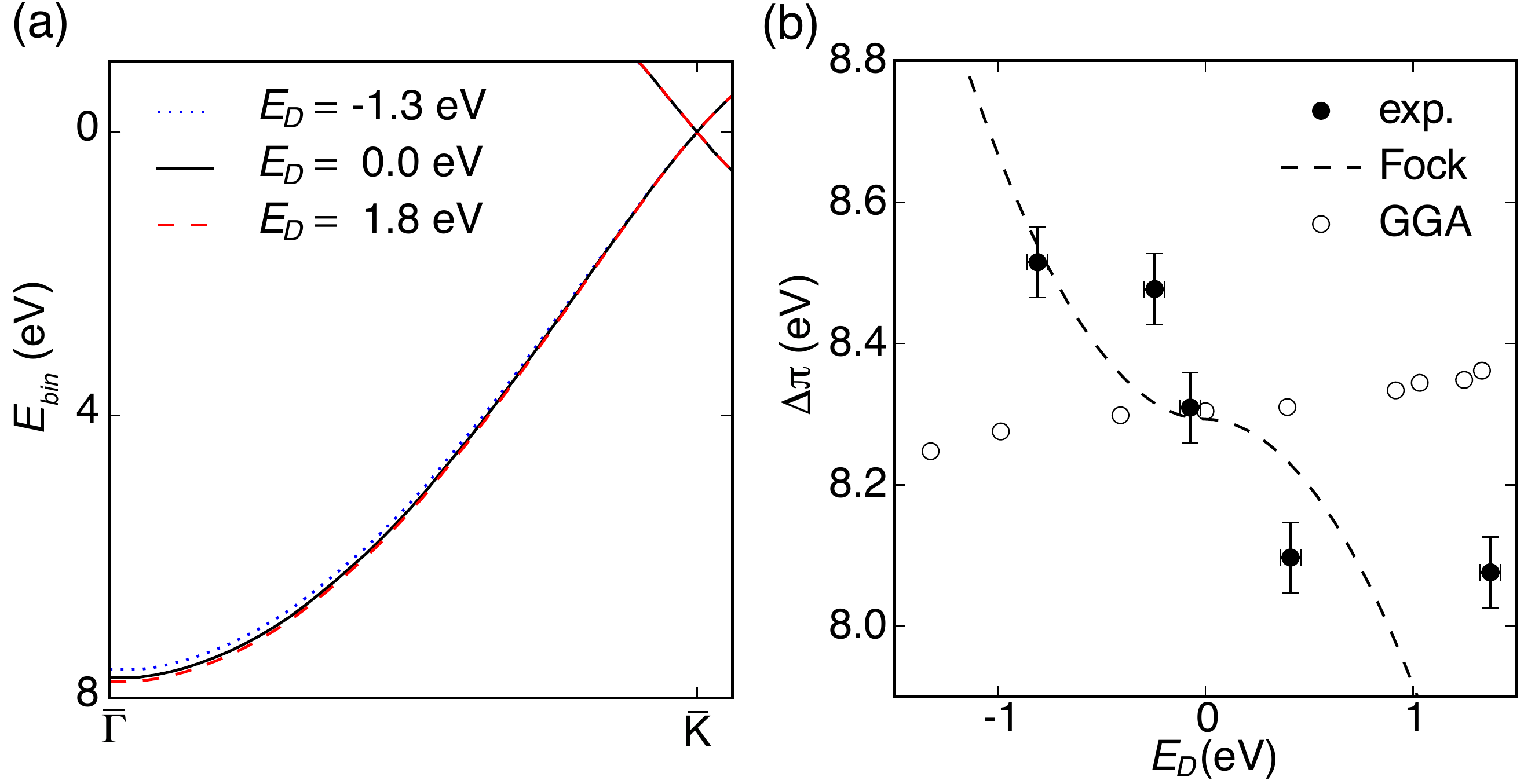}\\
  \caption{(Color online) (a) DFT band structure of graphene at different doping levels indicated by the Dirac point energies $E_D$ relative to the Fermi level. The band structures are aligned such that the Dirac point is at $E=0$ in all cases. (b) Bandwidth as function of the Dirac point energy considering Coulomb interaction in Hartree-Fock approximation (dashed line) compared to the experimental data and GGA results (constantly shifted by $+0.6$\,eV).}
  \label{fig:3}
\end{figure}

To trace the origin of the bandwidth change with doping we compare the experimental results to DFT calculations and estimates of the electronic self-energy based on perturbation theory. DFT calculations were performed with the Vienna Ab Initio Simulation Package (VASP) \cite{Kresse:1999} using the projector augmented wave (PAW) basis sets \cite{Blochl:1994,Kresse:1996}. Charge doping of graphene has been modelled by increasing / decreasing the number of electrons per unit cell and adding corresponding homogeneous compensating background charges. The resulting band structure of graphene at different doping levels as obtained from DFT calculations in the generalized gradient approximation (GGA) \cite{Perdew:1996} is shown in Fig. \ref{fig:3}(a). The GGA bandwidth of neutral graphene is $\Delta \pi=7.71$\,eV, which underestimates the measured band width by about $0.6$\,eV. The doping-induced band width changes are rather small on the scale of the full band width and can be more clearly seen in comparison to the measured $\Delta\pi$ in Fig. \ref{fig:3}(b). Here the GGA bandwidths are constantly shifted by $0.6$\,eV to match the experimental $\Delta\pi$ at the charge neutrality point for better comparability. We see that the GGA bandwidth $\Delta \pi$ increases upon electron doping, which is opposite to the experimental observation and suggests that many-body effects beyond GGA are responsible for the observed band width change. 

The effect of electronic interactions on electronic quasiparticles is encoded in the self-energy $\Sigma(k,\epsilon)$, which is in general a function of the wave-vector $k$ and the energy $\epsilon$. An extraction of band structures from semi-local approximations to DFT comes back to assuming that the self-energy is local (i.e., $k$-independent) and static (i.e., energy independent) and that it coincides with the exchange correlation potential. Given the failure of GGA to describe the experimentally observed bandwidth change, we search for the simplest non-local and / or dynamical self-energy terms which can be capable of explaining the experiments. 

The lowest order interaction effect yielding both kinds of contributions, is the screened exchange diagram. For a static non-local interaction $V_{ij}$ between two electrons at sites $i$ and $j$, the non-local Fock exchange term reads \[\Sigma_{ij}^F=-V_{ij}\langle c^\dagger_i c_j\rangle,\] where $c^\dagger_i$ ($c_j$) are the creation (annihilation) operators of the electrons and $\langle...\rangle$ denotes the quantum mechanical expectation value. We are thus interested in the change of $\langle c^\dagger_i c_j\rangle$ as function of the Fermi energy $\epsilon_f$:
\begin{equation}
\langle c^\dagger_i c_j\rangle=\sum_{k,\epsilon_k<\epsilon_f} \hpr{i}{k}\hpr{k}{j},
\label{eq:Fock1}
\end{equation}
where $\ket{k}$ denotes the single particle eigenstates of the electrons in graphene and $\epsilon_k$ their energy. For nearest-neighbor atoms, which belong to different sublattices, the phase factors in the integrand of Eq. (\ref{eq:Fock1}) turn out to wind around the Dirac point in the same way as the pseudospin vector of the Dirac fermions. Therefore, doping induced changes in $\langle c^\dagger_0 c_1\rangle$ cancel to lowest order when $\epsilon_f$ passes through the Dirac point and $\langle c^\dagger_0 c_1\rangle$ is an even function of $E_D-\epsilon_f$. Hence, nearest neighbor exchange scattering cannot be the origin of the observed band narrowing, which is an odd function of $E_D-\epsilon_f$.

However, next-nearest-neighbor atoms always belong to the same sublattice, which leads to $\langle c^\dagger_0 c_2\rangle$ being an odd function of $E_D-\epsilon_f$. An expansion of Eq. (\ref{eq:Fock1}) around the Dirac point yields $\langle c^\dagger_0 c_2\rangle=-n_e/4$, where $n_e$ is the doping level in electrons per unit cell, which we assume to be positive for electron doping and negative for hole doping. The resulting self-energy contribution, $\Sigma_{02}^F=V_{02}n_e/4$, renormalizes the next-nearest-neighbor hopping matrix element and thereby affects the band width according to $\delta(\Delta\pi)=\delta E_D - \delta E_\pi=-(9/4)V_{02}n_e$. I.e., the band width is reduced for electron doping and increased for hole doping, as observed experimentally. The exchange induced band width renormalizations as calculated numerically from Eq. \ref{eq:Fock1} is given as function of the Dirac point energy in Fig. \ref{fig:3}(b) for a next nearest neighbor interaction of  $V_{02}=4$\,eV. With this choice of the interaction we reproduce the experimental trend of band width changes qualitatively, while there are obvious quantitative differences between experiment and theory.

%\tw{*** Paragraph below could be used later in iterations of paper with referees *** would keep it out for now.}

%Ab-initio calculations of non-local Coulomb interactions in graphene \cite{Wehling_PRL11} yield values for the nearest-neighbor terms between $4$\,eV for free standing graphene and $\sim 0.5$\,eV for graphene encapsulated between two metallic Ir layers \cite{Roesner_WFCE_2015} if screening by the graphene $\pi$-electrons is disregarded. Screening of the $\pi$-electrons is expected to further reduce the interaction matrix elements. Therefore, a range of $V_{02}=2 - 4$\,eV as estimated from the comparison of the Fock exchange term and the experiment appears realistic as regards the order of magnitude, while the quantitative values appear at the upper bound of what should be expected from Ref. \cite{Roesner_WFCE_2015}. Hence, the Fock exchange is likely giving a significant but possibly not the only contribution to the experimentally observed band width change.

Possible further contributions can be substrate effects as well as dynamical (i.e, frequency dependent) terms entering the self-energy such as resulting from the coupling of the graphene $\pi$-electrons and bosons such as phonons. Phononic contributions to the self-energy turn out to lift the Dirac point up in energy for electron doping and lower its energy for hole doping on the typical scale of a few 10 meV \cite{Louie_Park_PRL07}. Therefore, phonon contributions to the dynamic part of the self-energy yield an effect which is too small in magnitude to explain the experimentally observed shifts and which would furthermore have the opposite sign to our observation.

In conclusion, the ARPES experiments reported here allow for the direct observation of doping-induced quasi particle band renormalizations. We find in particular deviations on the order of a several $100$\,meV of the measured band width from rigid band models and also from DFT GGA calculations, which is highly indicative of many-body effects. The simplest effect which is qualitatively able to explain these deviations is non-local exchange scattering. A full quantitative understanding of the observed band width renormalizations remains to be established. The experiments at hand can indeed define a benchmark case for the realiability of different electronic structure approaches.

We gratefully acknowledge funding from the VILLUM foundation, the Lundbeck foundation and the Danish Council for Independent Research, Natural Sciences under the Sapere Aude program (Grant Nos. DFF-4002-00029 and DFF-4090-00125), the DFG SPP 1459 and the European Graphene Flagship.  Work in Erlangen and Chemnitz was supported by the German Research Foundation (DFG).

%\bibliographystyle{apsrev}
%\bibliography{bandwidthbib}

\begin{thebibliography}{39}
\expandafter\ifx\csname natexlab\endcsname\relax\def\natexlab#1{#1}\fi
\expandafter\ifx\csname bibnamefont\endcsname\relax
  \def\bibnamefont#1{#1}\fi
\expandafter\ifx\csname bibfnamefont\endcsname\relax
  \def\bibfnamefont#1{#1}\fi
\expandafter\ifx\csname citenamefont\endcsname\relax
  \def\citenamefont#1{#1}\fi
\expandafter\ifx\csname url\endcsname\relax
  \def\url#1{\texttt{#1}}\fi
\expandafter\ifx\csname urlprefix\endcsname\relax\def\urlprefix{URL }\fi
\providecommand{\bibinfo}[2]{#2}
\providecommand{\eprint}[2][]{\url{#2}}

\bibitem[{\citenamefont{Luttinger}(1960)}]{Luttinger:1960}
\bibinfo{author}{\bibfnamefont{J.~M.} \bibnamefont{Luttinger}},
  \bibinfo{journal}{Physical Review} \textbf{\bibinfo{volume}{119}},
  \bibinfo{pages}{1153} (\bibinfo{year}{1960}).

\bibitem[{\citenamefont{Louie}(1992)}]{Louie:1992}
\bibinfo{author}{\bibfnamefont{S.~G.} \bibnamefont{Louie}}, in
  \emph{\bibinfo{booktitle}{Angle-resolved photoemission}}, edited by
  \bibinfo{editor}{\bibfnamefont{S.~D.} \bibnamefont{Kevan}}
  (\bibinfo{publisher}{Elsevier}, \bibinfo{address}{Amsterdam},
  \bibinfo{year}{1992}), vol.~\bibinfo{volume}{74} of
  \emph{\bibinfo{series}{Studies in Surface Chemistry and Catalysis}}.

\bibitem[{\citenamefont{Kalt and Rinker}(1992)}]{Kalt:1992aa}
\bibinfo{author}{\bibfnamefont{H.}~\bibnamefont{Kalt}} \bibnamefont{and}
  \bibinfo{author}{\bibfnamefont{M.}~\bibnamefont{Rinker}},
  \bibinfo{journal}{Phys. Rev. B} \textbf{\bibinfo{volume}{45}},
  \bibinfo{pages}{1139} (\bibinfo{year}{1992}).

\bibitem[{\citenamefont{Tr\"ankle et~al.}(1987)\citenamefont{Tr\"ankle, Leier,
  Forchel, Haug, Ell, and Weimann}}]{Trankle:1987aa}
\bibinfo{author}{\bibfnamefont{G.}~\bibnamefont{Tr\"ankle}},
  \bibinfo{author}{\bibfnamefont{H.}~\bibnamefont{Leier}},
  \bibinfo{author}{\bibfnamefont{A.}~\bibnamefont{Forchel}},
  \bibinfo{author}{\bibfnamefont{H.}~\bibnamefont{Haug}},
  \bibinfo{author}{\bibfnamefont{C.}~\bibnamefont{Ell}}, \bibnamefont{and}
  \bibinfo{author}{\bibfnamefont{G.}~\bibnamefont{Weimann}},
  \bibinfo{journal}{Phys. Rev. Lett.} \textbf{\bibinfo{volume}{58}},
  \bibinfo{pages}{419} (\bibinfo{year}{1987}).

\bibitem[{\citenamefont{Das~Sarma et~al.}(1990)\citenamefont{Das~Sarma,
  Jalabert, and Yang}}]{Das-Sarma:1990aa}
\bibinfo{author}{\bibfnamefont{S.}~\bibnamefont{Das~Sarma}},
  \bibinfo{author}{\bibfnamefont{R.}~\bibnamefont{Jalabert}}, \bibnamefont{and}
  \bibinfo{author}{\bibfnamefont{S.-R.~E.} \bibnamefont{Yang}},
  \bibinfo{journal}{Phys. Rev. B} \textbf{\bibinfo{volume}{41}},
  \bibinfo{pages}{8288} (\bibinfo{year}{1990}).

\bibitem[{\citenamefont{Jensen and Plummer}(1985)}]{Jensen:1985}
\bibinfo{author}{\bibfnamefont{E.}~\bibnamefont{Jensen}} \bibnamefont{and}
  \bibinfo{author}{\bibfnamefont{E.~W.} \bibnamefont{Plummer}},
  \bibinfo{journal}{Physical Review Letters} \textbf{\bibinfo{volume}{55}},
  \bibinfo{pages}{1912} (\bibinfo{year}{1985}).

\bibitem[{\citenamefont{{McClain} et~al.}(2015)\citenamefont{{McClain},
  Lischner, Watson, Matthews, Ronca, Louie, Berkelbach, and
  Chan}}]{mcclain_spectral_2015}
\bibinfo{author}{\bibfnamefont{J.}~\bibnamefont{{McClain}}},
  \bibinfo{author}{\bibfnamefont{J.}~\bibnamefont{Lischner}},
  \bibinfo{author}{\bibfnamefont{T.}~\bibnamefont{Watson}},
  \bibinfo{author}{\bibfnamefont{D.~A.} \bibnamefont{Matthews}},
  \bibinfo{author}{\bibfnamefont{E.}~\bibnamefont{Ronca}},
  \bibinfo{author}{\bibfnamefont{S.~G.} \bibnamefont{Louie}},
  \bibinfo{author}{\bibfnamefont{T.~C.} \bibnamefont{Berkelbach}},
  \bibnamefont{and} \bibinfo{author}{\bibfnamefont{G.~K.-L.}
  \bibnamefont{Chan}}, \bibinfo{journal}{ArXiv 1512.04556}
  (\bibinfo{year}{2015}), \eprint{1512.04556}.

\bibitem[{\citenamefont{Hedin}(1965)}]{Hedin:1965}
\bibinfo{author}{\bibfnamefont{L.}~\bibnamefont{Hedin}},
  \bibinfo{journal}{Phys. Rev.} \textbf{\bibinfo{volume}{139}},
  \bibinfo{pages}{A796} (\bibinfo{year}{1965}).

\bibitem[{\citenamefont{Northrup et~al.}(1989)\citenamefont{Northrup,
  Hybertsen, and Louie}}]{Northrup:1989}
\bibinfo{author}{\bibfnamefont{J.~E.} \bibnamefont{Northrup}},
  \bibinfo{author}{\bibfnamefont{M.~S.} \bibnamefont{Hybertsen}},
  \bibnamefont{and} \bibinfo{author}{\bibfnamefont{S.~G.} \bibnamefont{Louie}},
  \bibinfo{journal}{Phys. Rev. B} \textbf{\bibinfo{volume}{39}},
  \bibinfo{pages}{8198} (\bibinfo{year}{1989}).

\bibitem[{\citenamefont{Mahan and Sernelius}(1989)}]{Mahan:1989}
\bibinfo{author}{\bibfnamefont{G.~D.} \bibnamefont{Mahan}} \bibnamefont{and}
  \bibinfo{author}{\bibfnamefont{B.~E.} \bibnamefont{Sernelius}},
  \bibinfo{journal}{Phys. Rev. Lett.} \textbf{\bibinfo{volume}{62}},
  \bibinfo{pages}{2718} (\bibinfo{year}{1989}).

\bibitem[{\citenamefont{Zhu and Louie}(1991)}]{Zhu:1991}
\bibinfo{author}{\bibfnamefont{X.}~\bibnamefont{Zhu}} \bibnamefont{and}
  \bibinfo{author}{\bibfnamefont{S.~G.} \bibnamefont{Louie}},
  \bibinfo{journal}{Phys. Rev. B} \textbf{\bibinfo{volume}{43}},
  \bibinfo{pages}{14142} (\bibinfo{year}{1991}).

\bibitem[{\citenamefont{Mahan and Plummer}(2000)}]{Mahan:2000aa}
\bibinfo{author}{\bibfnamefont{G.~D.} \bibnamefont{Mahan}} \bibnamefont{and}
  \bibinfo{author}{\bibfnamefont{E.~W.} \bibnamefont{Plummer}}, in
  \emph{\bibinfo{booktitle}{Electronic Structure}}, edited by
  \bibinfo{editor}{\bibfnamefont{K.}~\bibnamefont{Horn}} \bibnamefont{and}
  \bibinfo{editor}{\bibfnamefont{M.}~\bibnamefont{Scheffler}}
  (\bibinfo{publisher}{Elsevier}, \bibinfo{address}{Amsterdam},
  \bibinfo{year}{2000}), Handbook of Surface Science.

\bibitem[{\citenamefont{He et~al.}(2015)\citenamefont{He, Hogan, Mion, Hafiz,
  He, Denlinger, Mo, Dhital, Chen, Lin et~al.}}]{He:2015ab}
\bibinfo{author}{\bibfnamefont{J.}~\bibnamefont{He}},
  \bibinfo{author}{\bibfnamefont{T.}~\bibnamefont{Hogan}},
  \bibinfo{author}{\bibfnamefont{T.~R.} \bibnamefont{Mion}},
  \bibinfo{author}{\bibfnamefont{H.}~\bibnamefont{Hafiz}},
  \bibinfo{author}{\bibfnamefont{Y.}~\bibnamefont{He}},
  \bibinfo{author}{\bibfnamefont{J.~D.} \bibnamefont{Denlinger}},
  \bibinfo{author}{\bibfnamefont{S.-K.} \bibnamefont{Mo}},
  \bibinfo{author}{\bibfnamefont{C.}~\bibnamefont{Dhital}},
  \bibinfo{author}{\bibfnamefont{X.}~\bibnamefont{Chen}},
  \bibinfo{author}{\bibfnamefont{Q.}~\bibnamefont{Lin}}, \bibnamefont{et~al.},
  \bibinfo{journal}{Nat Mater} \textbf{\bibinfo{volume}{14}},
  \bibinfo{pages}{577} (\bibinfo{year}{2015}).

\bibitem[{\citenamefont{Riley et~al.}(2015)\citenamefont{Riley, Meevasana,
  Bawden, Asakawa, Takayama, Eknapakul, Kim, Hoesch, Mo, Takagi
  et~al.}}]{Riley:2015aa}
\bibinfo{author}{\bibfnamefont{J.~M.} \bibnamefont{Riley}},
  \bibinfo{author}{\bibfnamefont{W.}~\bibnamefont{Meevasana}},
  \bibinfo{author}{\bibfnamefont{L.}~\bibnamefont{Bawden}},
  \bibinfo{author}{\bibfnamefont{M.}~\bibnamefont{Asakawa}},
  \bibinfo{author}{\bibfnamefont{T.}~\bibnamefont{Takayama}},
  \bibinfo{author}{\bibfnamefont{T.}~\bibnamefont{Eknapakul}},
  \bibinfo{author}{\bibfnamefont{T.~K.} \bibnamefont{Kim}},
  \bibinfo{author}{\bibfnamefont{M.}~\bibnamefont{Hoesch}},
  \bibinfo{author}{\bibfnamefont{S.~K.} \bibnamefont{Mo}},
  \bibinfo{author}{\bibfnamefont{H.}~\bibnamefont{Takagi}},
  \bibnamefont{et~al.}, \bibinfo{journal}{Nat Nano}
  \textbf{\bibinfo{volume}{10}}, \bibinfo{pages}{1043} (\bibinfo{year}{2015}).

\bibitem[{\citenamefont{Steinhoff et~al.}(2014)\citenamefont{Steinhoff,
  R{\"o}sner, Jahnke, Wehling, and Gies}}]{Steinhoff_NanoLett2014}
\bibinfo{author}{\bibfnamefont{A.}~\bibnamefont{Steinhoff}},
  \bibinfo{author}{\bibfnamefont{M.}~\bibnamefont{R{\"o}sner}},
  \bibinfo{author}{\bibfnamefont{F.}~\bibnamefont{Jahnke}},
  \bibinfo{author}{\bibfnamefont{T.~O.} \bibnamefont{Wehling}},
  \bibnamefont{and} \bibinfo{author}{\bibfnamefont{C.}~\bibnamefont{Gies}},
  \bibinfo{journal}{Nano Letters} \textbf{\bibinfo{volume}{14}},
  \bibinfo{pages}{3743} (\bibinfo{year}{2014}).

\bibitem[{\citenamefont{Larciprete et~al.}(2012)\citenamefont{Larciprete,
  Ulstrup, Lacovig, Dalmiglio, Bianchi, Mazzola, Hornek{\ae}r, Orlando,
  Baraldi, Hofmann et~al.}}]{Larciprete:2012}
\bibinfo{author}{\bibfnamefont{R.}~\bibnamefont{Larciprete}},
  \bibinfo{author}{\bibfnamefont{S.}~\bibnamefont{Ulstrup}},
  \bibinfo{author}{\bibfnamefont{P.}~\bibnamefont{Lacovig}},
  \bibinfo{author}{\bibfnamefont{M.}~\bibnamefont{Dalmiglio}},
  \bibinfo{author}{\bibfnamefont{M.}~\bibnamefont{Bianchi}},
  \bibinfo{author}{\bibfnamefont{F.}~\bibnamefont{Mazzola}},
  \bibinfo{author}{\bibfnamefont{L.}~\bibnamefont{Hornek{\ae}r}},
  \bibinfo{author}{\bibfnamefont{F.}~\bibnamefont{Orlando}},
  \bibinfo{author}{\bibfnamefont{A.}~\bibnamefont{Baraldi}},
  \bibinfo{author}{\bibfnamefont{P.}~\bibnamefont{Hofmann}},
  \bibnamefont{et~al.}, \bibinfo{journal}{ACS Nano}
  \textbf{\bibinfo{volume}{6}}, \bibinfo{pages}{9551} (\bibinfo{year}{2012}).

\bibitem[{\citenamefont{Bostwick et~al.}(2010)\citenamefont{Bostwick, Speck,
  Seyller, Horn, Polini, Asgari, MacDonald, and Rotenberg}}]{Bostwick:2010}
\bibinfo{author}{\bibfnamefont{A.}~\bibnamefont{Bostwick}},
  \bibinfo{author}{\bibfnamefont{F.}~\bibnamefont{Speck}},
  \bibinfo{author}{\bibfnamefont{T.}~\bibnamefont{Seyller}},
  \bibinfo{author}{\bibfnamefont{K.}~\bibnamefont{Horn}},
  \bibinfo{author}{\bibfnamefont{M.}~\bibnamefont{Polini}},
  \bibinfo{author}{\bibfnamefont{R.}~\bibnamefont{Asgari}},
  \bibinfo{author}{\bibfnamefont{A.~H.} \bibnamefont{MacDonald}},
  \bibnamefont{and}
  \bibinfo{author}{\bibfnamefont{E.}~\bibnamefont{Rotenberg}},
  \bibinfo{journal}{Science} \textbf{\bibinfo{volume}{328}},
  \bibinfo{pages}{999} (\bibinfo{year}{2010}).

\bibitem[{\citenamefont{Speck et~al.}(2011)\citenamefont{Speck, Jobst, Fromm,
  Ostler, Waldmann, Hundhausen, Weber, and Seyller}}]{Speck:2011}
\bibinfo{author}{\bibfnamefont{F.}~\bibnamefont{Speck}},
  \bibinfo{author}{\bibfnamefont{J.}~\bibnamefont{Jobst}},
  \bibinfo{author}{\bibfnamefont{F.}~\bibnamefont{Fromm}},
  \bibinfo{author}{\bibfnamefont{M.}~\bibnamefont{Ostler}},
  \bibinfo{author}{\bibfnamefont{D.}~\bibnamefont{Waldmann}},
  \bibinfo{author}{\bibfnamefont{M.}~\bibnamefont{Hundhausen}},
  \bibinfo{author}{\bibfnamefont{H.~B.} \bibnamefont{Weber}}, \bibnamefont{and}
  \bibinfo{author}{\bibfnamefont{T.}~\bibnamefont{Seyller}},
  \bibinfo{journal}{Applied Physics Letters} \textbf{\bibinfo{volume}{99}},
  \bibinfo{eid}{122106} (pages~\bibinfo{numpages}{3}) (\bibinfo{year}{2011}).

\bibitem[{\citenamefont{Johannsen et~al.}(2013)\citenamefont{Johannsen,
  Ulstrup, Bianchi, Hatch, Guan, Mazzola, Hornek{\ae}r, Fromm, Raidel, Seyller
  et~al.}}]{Johannsen:2013}
\bibinfo{author}{\bibfnamefont{J.~C.} \bibnamefont{Johannsen}},
  \bibinfo{author}{\bibfnamefont{S.~U.} \bibnamefont{Ulstrup}},
  \bibinfo{author}{\bibfnamefont{M.}~\bibnamefont{Bianchi}},
  \bibinfo{author}{\bibfnamefont{R.}~\bibnamefont{Hatch}},
  \bibinfo{author}{\bibfnamefont{D.}~\bibnamefont{Guan}},
  \bibinfo{author}{\bibfnamefont{F.}~\bibnamefont{Mazzola}},
  \bibinfo{author}{\bibfnamefont{L.}~\bibnamefont{Hornek{\ae}r}},
  \bibinfo{author}{\bibfnamefont{F.}~\bibnamefont{Fromm}},
  \bibinfo{author}{\bibfnamefont{C.}~\bibnamefont{Raidel}},
  \bibinfo{author}{\bibfnamefont{T.}~\bibnamefont{Seyller}},
  \bibnamefont{et~al.}, \bibinfo{journal}{Journal of Physics: Condensed Matter}
  \textbf{\bibinfo{volume}{25}}, \bibinfo{pages}{094001}
  (\bibinfo{year}{2013}).

\bibitem[{\citenamefont{Pletikosic et~al.}(2009)\citenamefont{Pletikosic,
  Kralj, Pervan, Brako, Coraux, N'Diaye, Busse, and Michely}}]{Pletikosic:2009}
\bibinfo{author}{\bibfnamefont{I.}~\bibnamefont{Pletikosic}},
  \bibinfo{author}{\bibfnamefont{M.}~\bibnamefont{Kralj}},
  \bibinfo{author}{\bibfnamefont{P.}~\bibnamefont{Pervan}},
  \bibinfo{author}{\bibfnamefont{R.}~\bibnamefont{Brako}},
  \bibinfo{author}{\bibfnamefont{J.}~\bibnamefont{Coraux}},
  \bibinfo{author}{\bibfnamefont{A.~T.} \bibnamefont{N'Diaye}},
  \bibinfo{author}{\bibfnamefont{C.}~\bibnamefont{Busse}}, \bibnamefont{and}
  \bibinfo{author}{\bibfnamefont{T.}~\bibnamefont{Michely}},
  \bibinfo{journal}{Physical Review Letters} \textbf{\bibinfo{volume}{102}},
  \bibinfo{eid}{056808} (\bibinfo{year}{2009}).

\bibitem[{\citenamefont{Bostwick
  et~al.}(2007{\natexlab{a}})\citenamefont{Bostwick, Ohta, Seyller, Horn, and
  Rotenberg}}]{Bostwick:2007}
\bibinfo{author}{\bibfnamefont{A.}~\bibnamefont{Bostwick}},
  \bibinfo{author}{\bibfnamefont{T.}~\bibnamefont{Ohta}},
  \bibinfo{author}{\bibfnamefont{T.}~\bibnamefont{Seyller}},
  \bibinfo{author}{\bibfnamefont{K.}~\bibnamefont{Horn}}, \bibnamefont{and}
  \bibinfo{author}{\bibfnamefont{E.}~\bibnamefont{Rotenberg}},
  \bibinfo{journal}{Nature Physics} \textbf{\bibinfo{volume}{3}},
  \bibinfo{pages}{36} (\bibinfo{year}{2007}{\natexlab{a}}).

\bibitem[{\citenamefont{Ulstrup et~al.}(2014)\citenamefont{Ulstrup, Andersen,
  Bianchi, Barreto, Hammer, Hornek{\ae}r, and Hofmann}}]{Ulstrup:2014e}
\bibinfo{author}{\bibfnamefont{S.}~\bibnamefont{Ulstrup}},
  \bibinfo{author}{\bibfnamefont{M.}~\bibnamefont{Andersen}},
  \bibinfo{author}{\bibfnamefont{M.}~\bibnamefont{Bianchi}},
  \bibinfo{author}{\bibfnamefont{L.}~\bibnamefont{Barreto}},
  \bibinfo{author}{\bibfnamefont{B.}~\bibnamefont{Hammer}},
  \bibinfo{author}{\bibfnamefont{L.}~\bibnamefont{Hornek{\ae}r}},
  \bibnamefont{and} \bibinfo{author}{\bibfnamefont{P.}~\bibnamefont{Hofmann}},
  \bibinfo{journal}{2D Materials} \textbf{\bibinfo{volume}{1}},
  \bibinfo{pages}{025002} (\bibinfo{year}{2014}).

\bibitem[{\citenamefont{Castro~Neto et~al.}(2009)\citenamefont{Castro~Neto,
  Guinea, Peres, Novoselov, and Geim}}]{Castro-Neto:2009}
\bibinfo{author}{\bibfnamefont{A.~H.} \bibnamefont{Castro~Neto}},
  \bibinfo{author}{\bibfnamefont{F.}~\bibnamefont{Guinea}},
  \bibinfo{author}{\bibfnamefont{N.~M.~R.} \bibnamefont{Peres}},
  \bibinfo{author}{\bibfnamefont{K.~S.} \bibnamefont{Novoselov}},
  \bibnamefont{and} \bibinfo{author}{\bibfnamefont{A.~K.} \bibnamefont{Geim}},
  \bibinfo{journal}{Rev. Mod. Phys.} \textbf{\bibinfo{volume}{81}},
  \bibinfo{pages}{109} (\bibinfo{year}{2009}).

\bibitem[{\citenamefont{Elias et~al.}(2011)\citenamefont{Elias, Gorbachev,
  Mayorov, Morozov, Zhukov, Blake, Ponomarenko, Grigorieva, Novoselov, Guinea
  et~al.}}]{Elias:2011}
\bibinfo{author}{\bibfnamefont{D.~C.} \bibnamefont{Elias}},
  \bibinfo{author}{\bibfnamefont{R.~V.} \bibnamefont{Gorbachev}},
  \bibinfo{author}{\bibfnamefont{A.~S.} \bibnamefont{Mayorov}},
  \bibinfo{author}{\bibfnamefont{S.~V.} \bibnamefont{Morozov}},
  \bibinfo{author}{\bibfnamefont{A.~A.} \bibnamefont{Zhukov}},
  \bibinfo{author}{\bibfnamefont{P.}~\bibnamefont{Blake}},
  \bibinfo{author}{\bibfnamefont{L.~A.} \bibnamefont{Ponomarenko}},
  \bibinfo{author}{\bibfnamefont{I.~V.} \bibnamefont{Grigorieva}},
  \bibinfo{author}{\bibfnamefont{K.~S.} \bibnamefont{Novoselov}},
  \bibinfo{author}{\bibfnamefont{F.}~\bibnamefont{Guinea}},
  \bibnamefont{et~al.}, \bibinfo{journal}{Nature Physics}
  \textbf{\bibinfo{volume}{7}}, \bibinfo{pages}{701} (\bibinfo{year}{2011}).

\bibitem[{\citenamefont{Siegel et~al.}(2011)\citenamefont{Siegel, Park, Hwang,
  Deslippe, Fedorov, Louie, and Lanzara}}]{Siegel:2011}
\bibinfo{author}{\bibfnamefont{D.~A.} \bibnamefont{Siegel}},
  \bibinfo{author}{\bibfnamefont{C.-H.} \bibnamefont{Park}},
  \bibinfo{author}{\bibfnamefont{C.}~\bibnamefont{Hwang}},
  \bibinfo{author}{\bibfnamefont{J.}~\bibnamefont{Deslippe}},
  \bibinfo{author}{\bibfnamefont{A.~V.} \bibnamefont{Fedorov}},
  \bibinfo{author}{\bibfnamefont{S.~G.} \bibnamefont{Louie}}, \bibnamefont{and}
  \bibinfo{author}{\bibfnamefont{A.}~\bibnamefont{Lanzara}},
  \bibinfo{journal}{Proceedings of the National Academy of Sciences}
  \textbf{\bibinfo{volume}{108}}, \bibinfo{pages}{11365}
  (\bibinfo{year}{2011}).

\bibitem[{\citenamefont{Kotov et~al.}(2012)\citenamefont{Kotov, Uchoa, Pereira,
  Guinea, and Castro~Neto}}]{Kotov:2012}
\bibinfo{author}{\bibfnamefont{V.~N.} \bibnamefont{Kotov}},
  \bibinfo{author}{\bibfnamefont{B.}~\bibnamefont{Uchoa}},
  \bibinfo{author}{\bibfnamefont{V.~M.} \bibnamefont{Pereira}},
  \bibinfo{author}{\bibfnamefont{F.}~\bibnamefont{Guinea}}, \bibnamefont{and}
  \bibinfo{author}{\bibfnamefont{A.~H.} \bibnamefont{Castro~Neto}},
  \bibinfo{journal}{Rev. Mod. Phys.} \textbf{\bibinfo{volume}{84}},
  \bibinfo{pages}{1067} (\bibinfo{year}{2012}).

\bibitem[{\citenamefont{Siegel et~al.}(2013)\citenamefont{Siegel, Regan,
  Fedorov, Zettl, and Lanzara}}]{Siegel:2013}
\bibinfo{author}{\bibfnamefont{D.~A.} \bibnamefont{Siegel}},
  \bibinfo{author}{\bibfnamefont{W.}~\bibnamefont{Regan}},
  \bibinfo{author}{\bibfnamefont{A.~V.} \bibnamefont{Fedorov}},
  \bibinfo{author}{\bibfnamefont{A.}~\bibnamefont{Zettl}}, \bibnamefont{and}
  \bibinfo{author}{\bibfnamefont{A.}~\bibnamefont{Lanzara}},
  \bibinfo{journal}{Phys. Rev. Lett.} \textbf{\bibinfo{volume}{110}},
  \bibinfo{pages}{146802} (\bibinfo{year}{2013}).

\bibitem[{\citenamefont{Speck et~al.}(2010)\citenamefont{Speck, Ostler,
  R\"ohrl, Jobst, Waldmann, Hundhausen, Ley, Weber, and
  Seyller}}]{Speck:2010aa}
\bibinfo{author}{\bibfnamefont{F.}~\bibnamefont{Speck}},
  \bibinfo{author}{\bibfnamefont{M.}~\bibnamefont{Ostler}},
  \bibinfo{author}{\bibfnamefont{J.}~\bibnamefont{R\"ohrl}},
  \bibinfo{author}{\bibfnamefont{J.}~\bibnamefont{Jobst}},
  \bibinfo{author}{\bibfnamefont{D.}~\bibnamefont{Waldmann}},
  \bibinfo{author}{\bibfnamefont{M.}~\bibnamefont{Hundhausen}},
  \bibinfo{author}{\bibfnamefont{L.}~\bibnamefont{Ley}},
  \bibinfo{author}{\bibfnamefont{H.~B.} \bibnamefont{Weber}}, \bibnamefont{and}
  \bibinfo{author}{\bibfnamefont{T.}~\bibnamefont{Seyller}},
  \bibinfo{journal}{Materials Science Forum}
  \textbf{\bibinfo{volume}{645-648}}, \bibinfo{pages}{629}
  (\bibinfo{year}{2010}).

\bibitem[{\citenamefont{Emtsev et~al.}(2009)\citenamefont{Emtsev, Bostwick,
  Horn, Jobst, Kellogg, Ley, McChesney, Ohta, Reshanov, Rohrl
  et~al.}}]{Emtsev:2009}
\bibinfo{author}{\bibfnamefont{K.~V.} \bibnamefont{Emtsev}},
  \bibinfo{author}{\bibfnamefont{A.}~\bibnamefont{Bostwick}},
  \bibinfo{author}{\bibfnamefont{K.}~\bibnamefont{Horn}},
  \bibinfo{author}{\bibfnamefont{J.}~\bibnamefont{Jobst}},
  \bibinfo{author}{\bibfnamefont{G.~L.} \bibnamefont{Kellogg}},
  \bibinfo{author}{\bibfnamefont{L.}~\bibnamefont{Ley}},
  \bibinfo{author}{\bibfnamefont{J.~L.} \bibnamefont{McChesney}},
  \bibinfo{author}{\bibfnamefont{T.}~\bibnamefont{Ohta}},
  \bibinfo{author}{\bibfnamefont{S.~A.} \bibnamefont{Reshanov}},
  \bibinfo{author}{\bibfnamefont{J.}~\bibnamefont{Rohrl}},
  \bibnamefont{et~al.}, \bibinfo{journal}{Nature Materials}
  \textbf{\bibinfo{volume}{8}}, \bibinfo{pages}{203} (\bibinfo{year}{2009}).

\bibitem[{\citenamefont{Ostler et~al.}(2010)\citenamefont{Ostler, Speck, Gick,
  and Seyller}}]{Ostler:2010}
\bibinfo{author}{\bibfnamefont{M.}~\bibnamefont{Ostler}},
  \bibinfo{author}{\bibfnamefont{F.}~\bibnamefont{Speck}},
  \bibinfo{author}{\bibfnamefont{M.}~\bibnamefont{Gick}}, \bibnamefont{and}
  \bibinfo{author}{\bibfnamefont{T.}~\bibnamefont{Seyller}},
  \bibinfo{journal}{physica status solidi (b)} \textbf{\bibinfo{volume}{247}},
  \bibinfo{pages}{2924} (\bibinfo{year}{2010}).

\bibitem[{\citenamefont{Hoffmann et~al.}(2004)\citenamefont{Hoffmann,
  S{\o}ndergaard, Schultz, Li, and Hofmann}}]{Hoffmann:2004}
\bibinfo{author}{\bibfnamefont{S.~V.} \bibnamefont{Hoffmann}},
  \bibinfo{author}{\bibfnamefont{C.}~\bibnamefont{S{\o}ndergaard}},
  \bibinfo{author}{\bibfnamefont{C.}~\bibnamefont{Schultz}},
  \bibinfo{author}{\bibfnamefont{Z.}~\bibnamefont{Li}}, \bibnamefont{and}
  \bibinfo{author}{\bibfnamefont{P.}~\bibnamefont{Hofmann}},
  \bibinfo{journal}{Nuclear Instruments and Methods in Physics Research, A}
  \textbf{\bibinfo{volume}{523}}, \bibinfo{pages}{441} (\bibinfo{year}{2004}).

\bibitem[{\citenamefont{Bostwick
  et~al.}(2007{\natexlab{b}})\citenamefont{Bostwick, Ohta, McChesney, Seyller,
  Horn, and Rotenberg}}]{Bostwick:2007b}
\bibinfo{author}{\bibfnamefont{A.}~\bibnamefont{Bostwick}},
  \bibinfo{author}{\bibfnamefont{T.}~\bibnamefont{Ohta}},
  \bibinfo{author}{\bibfnamefont{J.~L.} \bibnamefont{McChesney}},
  \bibinfo{author}{\bibfnamefont{T.}~\bibnamefont{Seyller}},
  \bibinfo{author}{\bibfnamefont{K.}~\bibnamefont{Horn}}, \bibnamefont{and}
  \bibinfo{author}{\bibfnamefont{E.}~\bibnamefont{Rotenberg}},
  \bibinfo{journal}{Solid State Communications} \textbf{\bibinfo{volume}{143}},
  \bibinfo{pages}{63} (\bibinfo{year}{2007}{\natexlab{b}}).

\bibitem[{\citenamefont{Trevisanutto et~al.}(2008)\citenamefont{Trevisanutto,
  Giorgetti, Reining, Ladisa, and Olevano}}]{Trevisanutto:2008}
\bibinfo{author}{\bibfnamefont{P.~E.} \bibnamefont{Trevisanutto}},
  \bibinfo{author}{\bibfnamefont{C.}~\bibnamefont{Giorgetti}},
  \bibinfo{author}{\bibfnamefont{L.}~\bibnamefont{Reining}},
  \bibinfo{author}{\bibfnamefont{M.}~\bibnamefont{Ladisa}}, \bibnamefont{and}
  \bibinfo{author}{\bibfnamefont{V.}~\bibnamefont{Olevano}},
  \bibinfo{journal}{Phys. Rev. Lett.} \textbf{\bibinfo{volume}{101}},
  \bibinfo{pages}{226405} (\bibinfo{year}{2008}).

\bibitem[{\citenamefont{Heske et~al.}(1999)\citenamefont{Heske, Treusch,
  Himpsel, Kakar, Terminello, Weyer, and Shirley}}]{Heske:1999}
\bibinfo{author}{\bibfnamefont{C.}~\bibnamefont{Heske}},
  \bibinfo{author}{\bibfnamefont{R.}~\bibnamefont{Treusch}},
  \bibinfo{author}{\bibfnamefont{F.~J.} \bibnamefont{Himpsel}},
  \bibinfo{author}{\bibfnamefont{S.}~\bibnamefont{Kakar}},
  \bibinfo{author}{\bibfnamefont{L.~J.} \bibnamefont{Terminello}},
  \bibinfo{author}{\bibfnamefont{H.~J.} \bibnamefont{Weyer}}, \bibnamefont{and}
  \bibinfo{author}{\bibfnamefont{E.~L.} \bibnamefont{Shirley}},
  \bibinfo{journal}{Phys. Rev. B} \textbf{\bibinfo{volume}{59}},
  \bibinfo{pages}{4680} (\bibinfo{year}{1999}).

\bibitem[{\citenamefont{Kresse and Joubert}(1999)}]{Kresse:1999}
\bibinfo{author}{\bibfnamefont{G.}~\bibnamefont{Kresse}} \bibnamefont{and}
  \bibinfo{author}{\bibfnamefont{D.}~\bibnamefont{Joubert}},
  \bibinfo{journal}{Phys. Rev. B} \textbf{\bibinfo{volume}{59}},
  \bibinfo{pages}{1758} (\bibinfo{year}{1999}).

\bibitem[{\citenamefont{Bl\"ochl}(1994)}]{Blochl:1994}
\bibinfo{author}{\bibfnamefont{P.~E.} \bibnamefont{Bl\"ochl}},
  \bibinfo{journal}{Phys. Rev. B} \textbf{\bibinfo{volume}{50}},
  \bibinfo{pages}{17953} (\bibinfo{year}{1994}).

\bibitem[{\citenamefont{Kresse and Furthm\"uller}(1996)}]{Kresse:1996}
\bibinfo{author}{\bibfnamefont{G.}~\bibnamefont{Kresse}} \bibnamefont{and}
  \bibinfo{author}{\bibfnamefont{J.}~\bibnamefont{Furthm\"uller}},
  \bibinfo{journal}{Phys. Rev. B} \textbf{\bibinfo{volume}{54}},
  \bibinfo{pages}{11169} (\bibinfo{year}{1996}).

\bibitem[{\citenamefont{Perdew et~al.}(1996)\citenamefont{Perdew, Burke, and
  Ernzerhof}}]{Perdew:1996}
\bibinfo{author}{\bibfnamefont{J.~P.} \bibnamefont{Perdew}},
  \bibinfo{author}{\bibfnamefont{K.}~\bibnamefont{Burke}}, \bibnamefont{and}
  \bibinfo{author}{\bibfnamefont{M.}~\bibnamefont{Ernzerhof}},
  \bibinfo{journal}{Phys. Rev. Lett.} \textbf{\bibinfo{volume}{77}},
  \bibinfo{pages}{3865} (\bibinfo{year}{1996}).

\bibitem[{\citenamefont{Park et~al.}(2007)\citenamefont{Park, Giustino, Cohen,
  and Louie}}]{Louie_Park_PRL07}
\bibinfo{author}{\bibfnamefont{C.-H.} \bibnamefont{Park}},
  \bibinfo{author}{\bibfnamefont{F.}~\bibnamefont{Giustino}},
  \bibinfo{author}{\bibfnamefont{M.~L.} \bibnamefont{Cohen}}, \bibnamefont{and}
  \bibinfo{author}{\bibfnamefont{S.~G.} \bibnamefont{Louie}},
  \bibinfo{journal}{Phys. Rev. Lett.} \textbf{\bibinfo{volume}{99}},
  \bibinfo{pages}{086804} (\bibinfo{year}{2007}).

\end{thebibliography}
\end{document}